\documentclass[journal=nalefd,manuscript=letter]{achemso}
\usepackage{amssymb}
\usepackage{amsmath}
\mciteErrorOnUnknownfalse

\author{Martin Saurbrey Bjergfelt}
\affiliation[University of Copenhagen]
{Center for Quantum Devices, Niels Bohr Institute, University of Copenhagen, 2100 Copenhagen, Denmark}

\author{Damon J. Carrad}
\affiliation[University of Copenhagen]
{Center for Quantum Devices, Niels Bohr Institute, University of Copenhagen, 2100 Copenhagen, Denmark}
\alsoaffiliation[DTUEnergy]{Department of Energy Conversion and Storage, Technical University of Denmark, 2800 Kongens Lyngby, Denmark}

\author{Thomas Kanne}
\affiliation[University of Copenhagen]
{Center for Quantum Devices, Niels Bohr Institute, University of Copenhagen, 2100 Copenhagen, Denmark}

\author{Erik Johnson}
\affiliation[University of Copenhagen]
{Center for Quantum Devices, Niels Bohr Institute, University of Copenhagen, 2100 Copenhagen, Denmark}
\alsoaffiliation[Technical University of Denmark]
{DTU Mechanical Engineering, Technical University of Denmark, 2800 Kongens Lyngby, Denmark}

\author{Elisabetta M. Fiordaliso}
\affiliation[Technical University of Denmark]
{DTU Nanolab, Technical University of Denmark, 2800 Kongens Lyngby, Denmark}

\author{Thomas Sand Jespersen}
\affiliation[University of Copenhagen]
{Center for Quantum Devices, Niels Bohr Institute, University of Copenhagen, 2100 Copenhagen, Denmark}
\alsoaffiliation[DTUEnergy]{Department of Energy Conversion and Storage, Technical University of Denmark, 2800 Kongens Lyngby, Denmark}

\author{Jesper Nygård}
\email{nygard@nbi.ku.dk}
\affiliation[University of Copenhagen]
{Center for Quantum Devices, Niels Bohr Institute, University of Copenhagen, 2100 Copenhagen, Denmark}

\title{Superconductivity and parity preservation in as-grown In islands on InAs nanowires}


\begin{document}

\begin{abstract}
We report \emph{in-situ} synthesis of crystalline indium islands on InAs nanowires grown by molecular beam epitaxy. Structural analysis by transmission electron microscopy showed that In crystals grew in a tetragonal body-centred crystal structure within two families of orientations relative to wurtzite InAs. The crystalline islands had lengths $<500$~nm and low-energy surfaces, suggesting that growth was driven mainly by surface energy minimization. Electrical transport through In/InAs devices exhibited Cooper pair charging, evidencing charge parity preservation and a pristine In/InAs interface, with an induced superconducting gap $\sim 0.45$~meV. Cooper pair charging persisted to temperatures $>1.2$~K and magnetic fields $\sim 0.7$~T, demonstrating that In/InAs hybrids belong to an expanding class of semiconductor/superconductor hybrids operating over a wider parameter space than state-of-the-art Al-based hybrids. Engineering crystal morphology while isolating single islands using shadow epitaxy provides an interesting alternative to previous semiconductor/superconductor hybrid morphologies and device geometries.
\end{abstract}

\newpage

The prediction that semiconductor/superconductor hybrids may act as hosts of topological superconductivity has prompted a concerted effort to refine these nanostructures and extend their capabilities.\cite{Lutchyn2018a} In this respect, the development of epitaxially matched metal/semiconductor heterostructures\cite{Krogstrup2015, Chang2015} was a crucial step; hybrids with pristine interfaces exhibit a low density of states within the induced superconducting gap\cite{Krogstrup2015,Gazibegovic2017, Chang2015, Carrad2020,Pendharkar2021, Kanne2020} and enable the preservation of charge parity in semiconducting/superconducting hybrid islands.\cite{AlbrechtNature16,ShenNatComm18, Carrad2020, Kanne2020, Pendharkar2021} These advances highlight the benefits of utilising `bottom-up' approaches in nanodevice engineering and encourage the use of metal/semiconductor epitaxy of other materials and in new contexts.\cite{Gusken2017,Carrad2020, Bjergfelt2019c,Heedt2020, Kanne2020, Pendharkar2021} New materials can increase the available parameter space in terms of critical temperature, $T_\mathrm{C}$, and critical magnetic field, $B_\mathrm{C}$, compared to start-of-the-art Al-based hybrids\cite{Lutchyn2018a,Heedt2020,Gusken2017} as has been shown for Sn\cite{Pendharkar2021}, Pb\cite{Kanne2020}, Ta\cite{Carrad2020}, and V\cite{Carrad2020, Bjergfelt2019c}.  More broadly, it is also relevant to consider other structures beyond the common gate-tunable Josephson junctions\cite{DohScience05, Larsen2015} and tunnel probes\cite{Lee2014, Mourik2012b, Chang2015} by harnessing different materials and/or growth regimes, which may enable new functionalities and applications. In particular, the nanoscale island morphology has been exploited to extend fundamental understanding of, e.g., superconductivity\cite{Ruby2015} and magnetism,\cite{Oka2014, Gambardella2003} and signatures of topological superconductivity have been observed.\cite{Nadj-Perge2014, Menard2019} Further, the combination of bottom-up grown nanowires and quantum dots holds promise within quantum photonics.\cite{Uccelli2010,Heiss2013,Herranz2020} Here, we demonstrate a new type of hybrid, In/InAs, where the underlying growth dynamics caused In to form in islands along an InAs nanowire (NW). Seeking conditions for coherent thin film growth of In on InAs, we investigated different In deposition temperatures, rates, and techniques. Using shadow epitaxy\cite{Carrad2020} to isolate single In islands provided a path towards utilizing as-grown superconducting In islands on bottom-up grown InAs NWs. Electron transport at cryogenic temperatures revealed gate-controlled Cooper pair charging of single islands at temperatures up to 1.2~K and magnetic fields up to 0.7~T. The measurements confirm that In provides strong suppression of single electron tunneling, crucial for the operation of superconducting and topological qubits.\cite{VanWoerkom2015, Pendharkar2021,Arute2019,Koch2007,Bouchiat1998, Lutchyn2018a} So far, Cooper pair charging has only been demonstrated in a handful of materials.\cite{HergenrotherPRL94, TuominenPRL92,VanWoerkom2015, AlbrechtNature16, Kanne2020,Pendharkar2021} The presented path towards in-situ formed devices complements other techniques with a similar goal of obtaining and preserving pristine surfaces and interfaces by exploiting the benefits of bottom-up nanostructure definition.\cite{Gazibegovic2017,Carrad2020,Heedt2020}

\begin{figure}[!t]
\centering
\includegraphics[width=9cm]{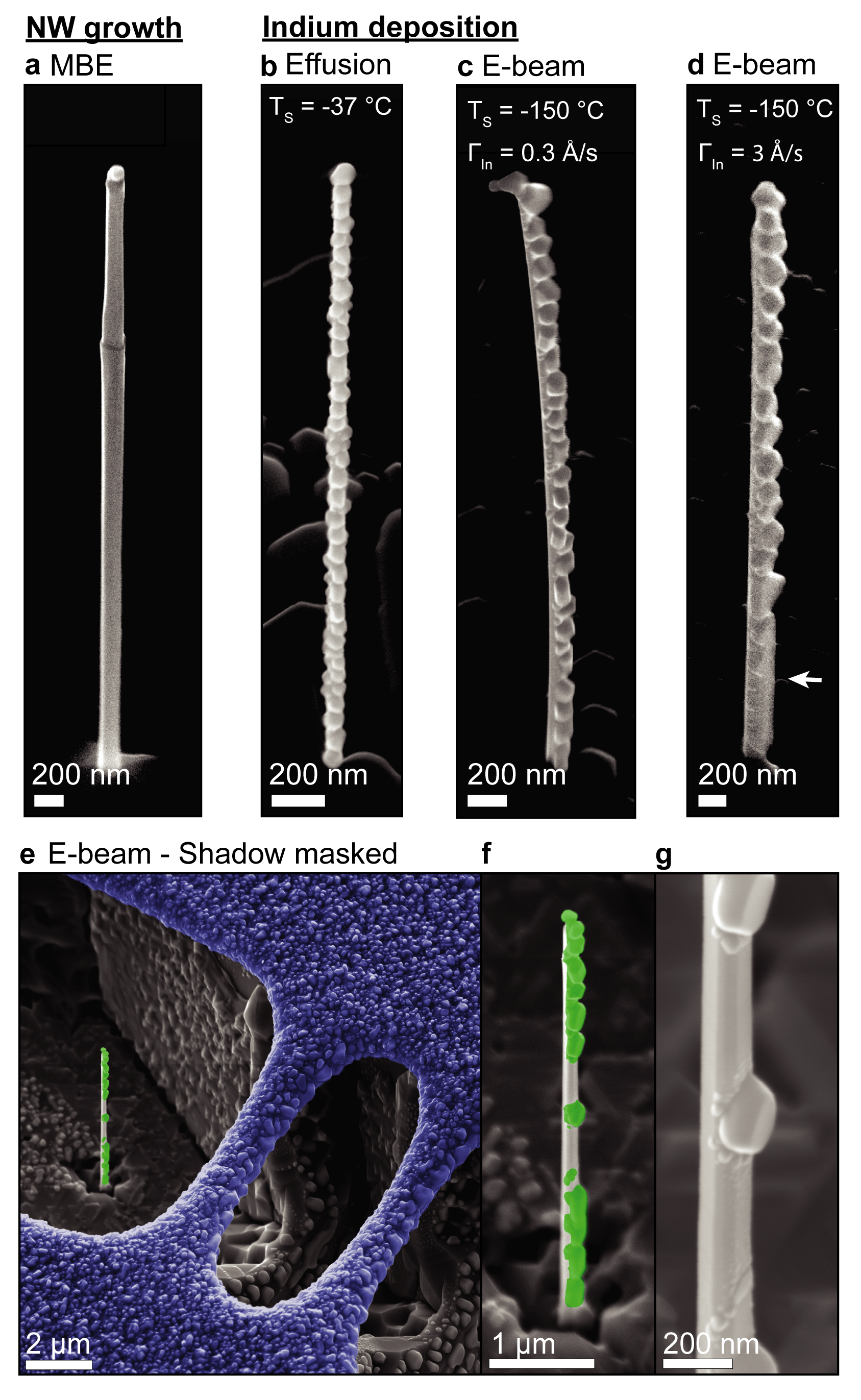}
\caption{\label{fig:In_1}(a) As-grown InAs wurtzite NW. (b-d) Representative NWs from growths 1-3, respectively, with varied material source, substrate temperature, $T_\mathrm{S}$, and In deposition rate, $\Gamma_\mathrm{In}$. (e-f) False-colored SEMs of as-grown hybrid NW growth substrate where the In (green) was evaporated onto an InAs NW through a SiOx shadow mask (blue) with a 10$^\circ$ angle between substrate surface and the crucible normal. The `dual bridge' shadow mask geometry resulted in two $\sim 600$~nm-long gaps in the In coverage.~\cite{Carrad2020} (f,g) Increased magnification reveals small -- $\sim 20$~nm diameter -- grains around the larger -- 50-200~nm diameter -- grains.}
\end{figure}

Figure~\ref{fig:In_1}a shows an InAs NW grown by molecular beam epitaxy (MBE) under conditions which produce well-defined NW facets (see Methods and Refs~\citenum{Carrad2020,Kanne2020}). We report here results from three growths, using two In deposition methods with different lower bounds on deposition temperature. For growth 1, we utilised MBE effusion cell deposition after NW growth, with substrate temperature $T_\mathrm{S}=-37^\circ$C\cite{Krogstrup2015} (Fig.~\ref{fig:In_1}b). An alternative method is to transfer the as-grown NWs to a separate chamber under UHV and use e-beam deposition\cite{Gusken2017,Bjergfelt2019c, Carrad2020}, where we achieve substrate temperatures down to $T_\mathrm{S}=-150^\circ$C. We performed two growths utilising the latter method, and tested two different deposition rates, $\Gamma_\mathrm{In}=0.3$ Å/s and $\Gamma_\mathrm{In}=3.0$ Å/s, for growths 2 and 3, respectively (Figures ~\ref{fig:In_1}c,d). The higher rate of deposition for growth 3 presumably promoted a higher density of In nucleation sites for the thin film growth~\cite{Ratsch2003}. Growths 2 and 3 had a nominal In thickness of 40~nm and were performed on shadow epitaxy wafers, where superconductor flux may be selectively shadowed to produce an array of different device geometries~\cite{Carrad2020}. A tilt angle $\theta= 10^\circ$ between the In flux and $[11\bar{2}]$ sample plane direction enabled the masks to cast a shadow on the NWs, while ensuring a 3-facet deposition. This concept is demonstrated in Figs.~1e-g, where two SiOx `bridges' (blue false color) generated two shadowed regions on the NW without In.

\begin{figure}[!t]
\centering
\includegraphics[width=\textwidth]{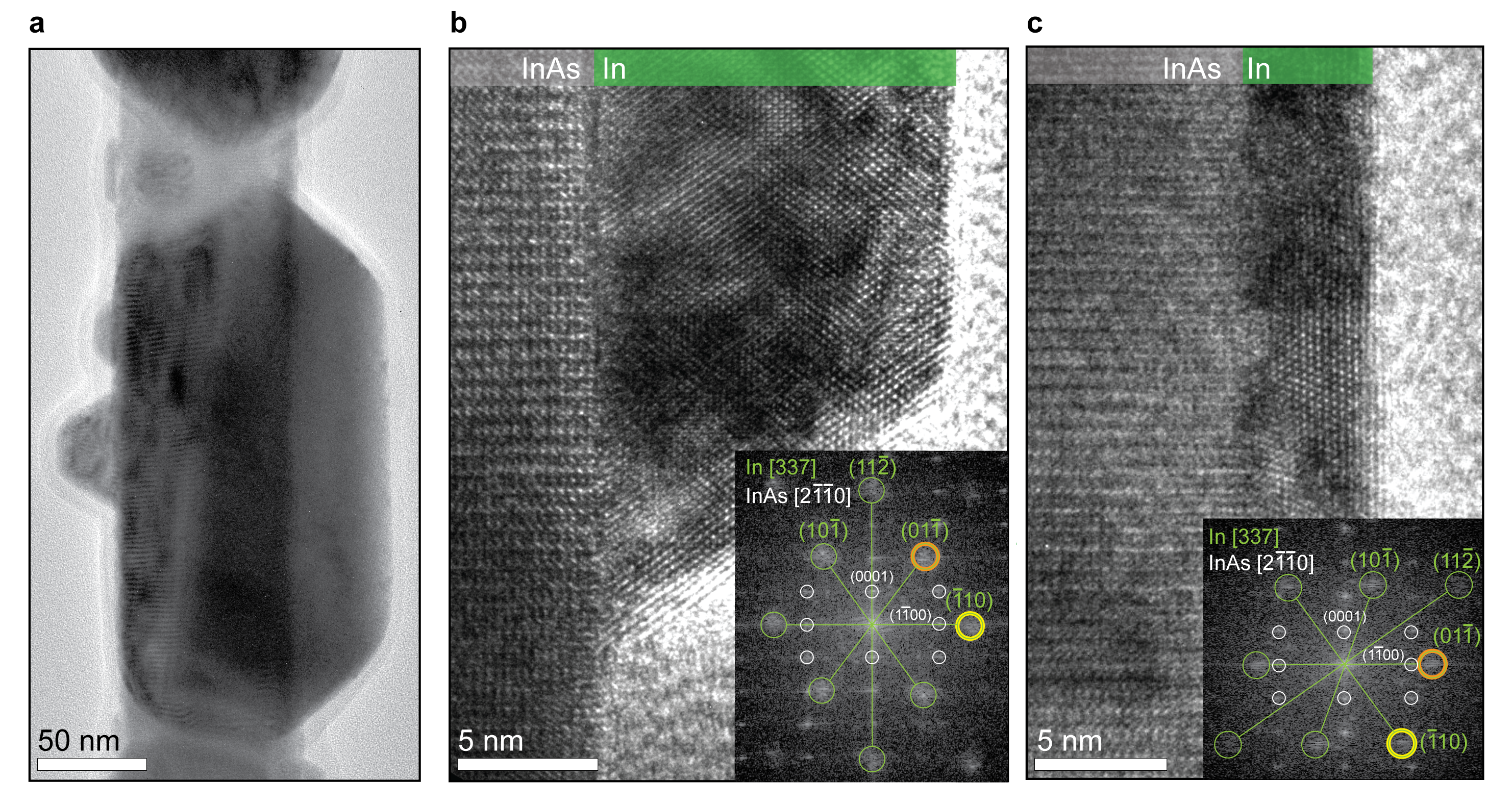}
\caption{\label{fig:In_2}Transmission electron microscopy (TEM) of three different In/InAs hybrids from growth 2. (a) TEM of an In/InAs NW with the electron beam aligned parallel to $\{1\bar{1}00\}$ facets. Faceted In grains of various sizes formed on the NW with In facets of each grain aligned to the underlying InAs surface facet. Moir\'e fringes are visible in overlapping grains, suggesting rotational variation. (b,c) HRTEMs of two selected In/InAs interfaces. Atomic lattice fringes are visible throughout the grains from In [337] zone-axes. FFT in the insert to (b) shows that $\lbrace \bar{1}10\rbrace$ and  $\lbrace 11\bar{2}\rbrace$ crystal planes in the In grain aligned with facet $(1\bar{1}00)$ and axial (0001) planes, respectively. FFT in the insert to (c) shows that $\lbrace 01\bar{1}\rbrace$ crystal planes in the In grain are aligned with InAs facet $(1\bar{1}00)$ planes.}
\end{figure}

\begin{figure}[!ht]
\centering
\includegraphics[width=15cm]{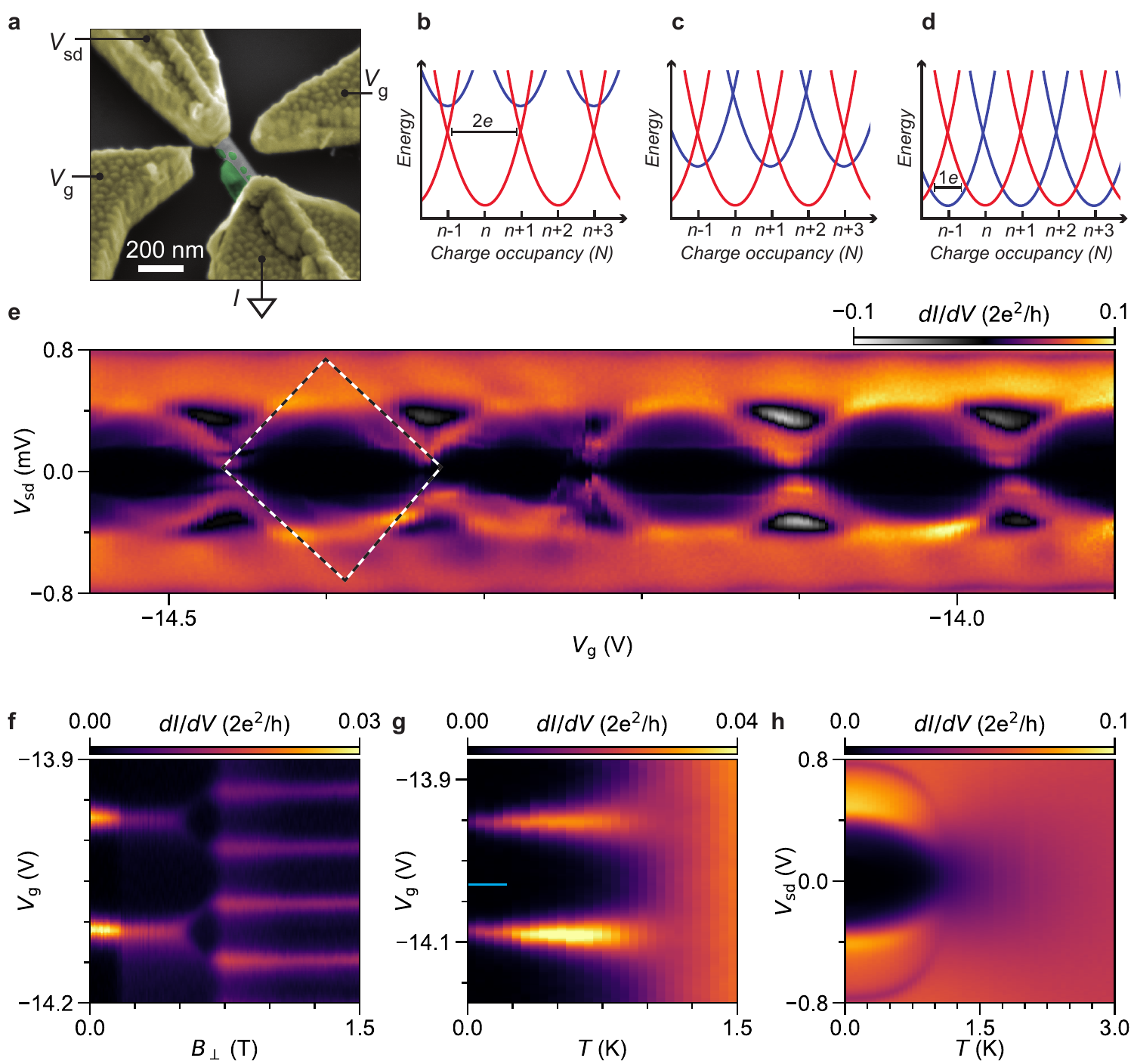}
\caption{\label{fig:In_3} (a) False colour SEM of In-coated (green) InAs NW (grey) contacted by two Ti/Au source/drain electrodes (gold) aligned to a single grain on the NW. Two nearby side-gate electrodes (gold) were used to tune the island charge occupation, NW charge density, and thus coupling between source/drain and the island. (b) $2e$-periodic transport arises due to odd occupation (blue parabolas) requiring additional energy, $\Delta$, compared to even occupation (red parabolas).\cite{TuominenPRL92} (c) Increased magnetic field, or the presence of a bound state, lowers the energy required for odd charge occupation. (d) $1e$-periodic transport occurs for $B > B_\mathrm{C}$ or a bound state at zero energy.~\cite{TuominenPRL92, AlbrechtNature16}. (e) Differential conductance, $dI/dV$, versus source-drain bias, $V_{\mathrm{sd}}$, and gate voltage $V_\mathrm{g}$ shows a 2$e$-periodic diamond structure characteristic for superconducting islands~\cite{AlbrechtNature16,HergenrotherPRL94} with $\Delta \sim 0.45$ meV. Dashed lines are a guide outlining the border of a single Coulomb diamond. (f) Applying perpendicular magnetic field to the In hybrid broke 2$e$-periodic charging into 1$e$-periodic charging at $B_{\perp} \sim 0.7$~T. (g) The island remained 2$e$-periodic up to at least $T = 1.2$ K. (h) $dI/dV$ vs $V_\mathrm{sd}$ vs $T$ for fixed $V_\mathrm{g}=-14.03$~V (blue line in (g)).}
\end{figure}

Considering first broad structural features, the scanning electron micrographs (SEMs) in Figs.~1b-d show that In exhibited a granular morphology across all growths, despite the variation in $T_\mathrm{S}$ and $\Gamma_\mathrm{In}$. The grain lengths and thicknesses ranged from $\sim 10 - 500$~nm and $\sim 5 - 40$~nm, respectively with the smallest grains most clearly visible in Fig.~\ref{fig:In_1}g, and a particularly long grain visible at the base of the NW in Fig.~\ref{fig:In_1}d marked with an arrow. The low melting temperature of In ($T_m=157^\circ$C),\cite{Haynes2016} high probability for dewetting and imply a large diffusion constant for In adatoms on crystalline surfaces, such as InAs.\cite{Mehrer2007,Thompson2012,Kanne2020} A large diffusion constant can cause a lower density of stable clusters and thus a relatively larger distance between the adsorbed clusters, which may, in turn, grow as individual Volmer-Weber islands.\cite{Pimpinelli1998} Lowering the substrate temperature and increasing the flux during growth has been shown to increase the density of nucleation sites.\cite{Pentcheva2003} This is consistent with the observation of the longest In segment in Fig.~\ref{fig:In_3}d (white arrow), with $T_\mathrm{S}=150^\circ$C, $\Gamma_\mathrm{In} = 3$~Å/s.

Figure~\ref{fig:In_2} shows transmission electron microscope (TEM) micrographs of selected NWs from growth 2. The In grains in Fig.~\ref{fig:In_2}a exhibited well-defined facets, suggestive of an underlying crystalline order. High resolution TEM (HRTEM) and the associated fast Fourier transforms (FFTs) of individual grains (Figs.~2b,c) confirmed that In crystallised in the tetragonal body-centred (TBC) space group $I4/mmm$, as in bulk.~\cite{Moshopoulou2006,Haynes2016} We established this by comparing the interplanar distances and angles between planes measured from Figs.~2b,c to literature values for various bulk In TBC planes (see supplementary section for structural analysis). The assigned planes are given in Figs.~2b,c insets. For all 17 grains studied by HRTEM from growths 2 and 3, a $\langle 337\rangle$ In zone-axis was aligned with the $\langle2\bar{1}\bar{1}0\rangle$ InAs zone-axis in both growths. Different In crystal alignments to the NW facet out-of-plane ($[1\bar{1}00]$) and axial plane ([0001]) orientations were observed among the studied grains; twelve grains featured a [110] out-of-plane and [$11\bar{2}$] in-plane alignment (Fig. \ref{fig:In_3}b), three grains featured a [101] out-of-plane alignment and no in-plane alignment (Fig. \ref{fig:In_2}c), and two grains did not feature any out-of-plane alignment to the NW. The fact that most In grains formed with low-index planes parallel to the InAs ($1\bar{1}00$) surface facets is supported by dark field TEM imaging of 8 different grains in the same NW region featuring these out-out-plane orientations (Supplementary Fig. S1). With the $\langle 337 \rangle $ zone-axes and the observed out-plane orientations, 3 sets of 8 symmetrically equal In orientations on InAs are thus possible. From these 24 orientations, our observations suggest that the 8 variations featuring [110] out-of-plane and [$11\bar{2}$] in-plane alignments have a higher probability of forming.

To explain the morphology of In on InAs, we consider the competition between thermodynamic driving forces -- consisting of surface, interface and strain energies -- and kinetic limiting factors -- including adatom diffusion and cluster formation.~\cite{Thompson1990,Ratsch2003,Kanne2020,Krogstrup2015}. For In on InAs, the observation of $\{110\}$ and $\{101\}$ planes parallel to the InAs surface in all grains, but not the $\{11\bar{2}\}$ planes -- which also belong to the $\langle 337\rangle$ zone axis family -- indicates that grain formation during growth was mainly driven by minimisation of grain surface energies;  $\{110\}$ and $\{101\}$ surfaces are energetically favoured compared to $\{11\bar{2}\}$.\cite{Thompson1990,Tran2016} Similarly, the fact that each In island consisted of a single crystal indicates that the energetic cost of recrystallization may have been lower than the cost of forming grain boundaries. Formation of a multitude of disconnected single crystalline grains with a large range of sizes suggests that the probability for the In to form a nucleation site was low or that the critical nucleation size for forming a stable cluster was not reached. In this scenario, only a small number of grains would nucleate and grow compared to the volume of material from the incoming In flux, resulting in the observed partial coverage. Overall, the observed characteristics of single crystal, disconnected grains were likely driven by a minimisation of grain surface energies, and consistent with previous theoretical predictions regarding growth of In on InAs (see supplementary information of Ref. 7).

Having obtained a new class of hybrid in terms of materials, we next characterise its behaviour using electron transport. Superconducting islands play an important role in technologically relevant devices; the charge parity of superconducting islands defines the fundamental behaviour of state-of-the-art superconducting qubits\cite{Arute2019,Koch2007,Bouchiat1998}, and hybrid semiconductor/superconductor islands are a leading candidate as building blocks for topologically protected quantum computation.~\cite{Lutchyn2018a} The superconducting properties of a single In grain were studied using electron transport through the In/InAs device shown in Fig.~\ref{fig:In_3}a. The NW was contacted by two Ti/Au source/drain electrodes aligned to a single grain on the NW. Two Ti/Au side-gates with gate voltage $V_{\mathrm{g}}$ tuned the charge occupation of the In island, the NW charge density, and thereby also the coupling between the source/drain and island. As a result, $V_\mathrm{g}$ was used to tune the current, $I$, passing through the device in response to applied source-drain voltage, $V_\mathrm{sd}$. At negative $V_\mathrm{g}$, the coupling was weak and the device was in Coulomb blockade. Coulomb diamonds with 2$e$-periodicity were observed in differential conductance, $dI/dV$, when measured as a function of $V_\mathrm{g}$ (Fig.~\ref{fig:In_3}e), characteristic of a superconducting island, where the superconducting segment is weakly coupled to both leads, and develops a charging energy $E_\mathrm{C}$.~\cite{TuominenPRL92, Lafarge1993, HergenrotherPRL94, HigginbothamNatPhys15, AlbrechtNature16, ShenNatComm18, Carrad2020, Kanne2020} The top/bottom diamond tips corresponds to $eV_\mathrm{sd} = \pm 8E_\mathrm{C} \sim 0.6$~meV, giving $E_\mathrm{C}=75~\mu$eV. The regions of negative differential conductance result from the onset of quasiparticle poisoning at $V_\mathrm{sd} = \Delta/e$,\cite{HergenrotherPRL94, AlbrechtNature16, ShenNatComm18, Kanne2020} giving $\Delta = 0.45$~meV for this device. Deviation from the bulk value, $\Delta_{\mathrm{bulk}} = 0.53$~meV,~\cite{Kittel2004} in this nanostructured hybrid may have arisen from e.g. size-related effects and/or hybridisation due to the proximity effect.\cite{Bose2014,Li2005,Reeg2017}

Transport through Coulomb blockaded islands around $V_\mathrm{sd} = 0$ occurs by tuning $V_\mathrm{g}$ to the periodic degeneracy points between the parabolic energy dispersion relations of each charge occupancy level, $N$.\cite{TuominenPRL92, Lafarge1993} Fig.~\ref{fig:In_3}b shows that, for superconducting islands at low temperature and zero magnetic field, the energy required for the island to host an odd number of electrons (blue parabolas) compared to an even number (red parabolas) is increased by $\Delta$. Consequently, for $E_\mathrm{C}\leq \Delta$, Coulomb peaks occur with $2e$-periodicity and demarcate regions in $V_\mathrm{g}$ of successive even-occupied charge states. With increased magnetic field (Fig.~\ref{fig:In_3}c), $\Delta$ falls below the $2e$ degeneracy points, permitting odd electron occupation and doubling the number of peaks. Finally, single electron tunnelling with $1e$-periodicity occurs for $\Delta = 0$, (Fig.~\ref{fig:In_3}d).~\cite{LuPRB96} These behaviours are observed in Fig.~\ref{fig:In_3}f, where $2e$-periodic charging at zero perpendicular magnetic field, $B_\perp = 0$ evolves into $1e$-periodic charging above a characteristic field $B_\mathrm{C} \sim 0.7$~T. The $1e$-periodicity above $B_\perp = 0.7$~T may have resulted from $B_\perp$ approaching the critical magnetic field for this sample,~\cite{LuPRB96} or a bound state below the superconducting energy gap converging to zero energy.\cite{HigginbothamNatPhys15, AlbrechtNature16, VaitiekenasPRL18sag, Carrad2020, Kanne2020, ShenNatComm18, Vaitiekenasfullshell18,Pendharkar2021} While the perpendicular field orientation and absence of field-dependent peak spacing oscillations discounts the possibility of a topologically non-trivial state emerging for $B_\perp > 0.7$~T, the observations of $2e$-periodicity and a substantial critical field -- lower bound $0.7$~T -- highlight the potential for In/InAs hybrids to host Majorana modes.~\cite{AlbrechtNature16, Vaitiekenasfullshell18, ShenNatComm18} Additionally, the increase in perpendicular critical field over that for Al suggests In may find use in fundamental studies or applications in superconducting qubits.\cite{AlbrechtNature16,Larsen2015}

Importantly, $2e$-periodic transport is not universal in superconducting materials and has previously been reported only for Al\cite{TuominenPRL92, Lafarge1993} and NbTiN islands\cite{VanWoerkom2015}, and hybrid devices based on Al,~\cite{AlbrechtNature16, VaitiekenasPRL18sag, Carrad2020, ShenNatComm18} Pb~\cite{Kanne2020} and Sn.~\cite{Pendharkar2021} The fact that In/InAs islands support tunneling of Cooper pairs and charge parity preservation implies high material quality, including a hard superconducting gap. The presence of impurity-related sub-gap states in soft-gap hybrids results in zero-field $1e$-periodic charging,~\cite{AlbrechtNature16, Kanne2020, ShenNatComm18,Pendharkar2021} and leads to decoherence in superconducting qubits.\cite{VanWoerkom2015,Bouchiat1998,Koch2007} Figure 3g shows that the $2e$-periodic charging persists with elevated temperatures up to $T \sim 1$~K, where the Coulomb blockade peaks begin to overlap significantly. This temperature is consistent with $k_\mathrm{B}T$ exceeding the charging energy $E_\mathrm{C} = 75~\mu$eV\cite{Beenakker1991}, and we attribute the loss of $2e$-periodicity to the loss of Coulomb blockade. In Fig. \ref{fig:In_3}h, we show temperature-dependent tunneling spectroscopy, with $V_\mathrm{g}$ fixed between two degeneracy points (blue line in Fig.~\ref{fig:In_3}g). At $T=0$, $dI/dV$ is suppressed around $V_\mathrm{sd}=0$ due to the dual effects of Coulomb blockade and the superconducting gap. At $T\sim 1$~K, zero bias conduction is enhanced by the loss of Coulomb blockade, with a suppression around zero bias remaining until $T\sim 3$~K, consistent with the hybrid $T_\mathrm{C}$. The temperature-dependent behaviour of indium contrasts the temperature-induced transitions from $2e$ to $1e$ charging in Al islands at $T \sim 200-300$~mK  which have been attributed to thermally activated quasi-particle poisoning dominating transport.\cite{TuominenPRL92, HigginbothamNatPhys15, VaitiekenasPRL18sag, Carrad2020,Lafarge1993} The temperature-dependent charging in Figure~\ref{fig:In_3}g shows that in the In/InAs device, the quasi-particle poisoning temperature is at least three times larger than previously observed for Al-based devices. This is the first demonstration of a quasi-particle-free superconducting island operating at Kelvin temperatures, which encourages further the development of temperature resilient semiconductor/superconductor hybrids~\cite{Kanne2020,Pendharkar2021} and superconducting qubits.\cite{Arute2019,Koch2007,Bouchiat1998}

In summary, indium formed disconnected crystalline grains on InAs NWs. The grains formed in the bulk TBC phase with well defined low-index surfaces suggested that the grain formations and re-crystallisations primarily were driven by surface energy minimisation. Electrically, the grains behaved as superconducting islands. An increased poisoning temperature combined with the fact that the superconducting gap remained close to the bulk value, suggests In as a potentially useful material assuming ultimate control of film/island growth. Controlling island size would be an interesting route to obtain chains of superconducting islands for, e.g., Kitaev chains~\cite{Kitaev2001}, or to study Majorana interactions between two adjacent islands~\cite{Hassler2012,AlbrechtNature16,Vaitiekenasfullshell18}. Achieving a continuous thin film would allow for further characterisation of the induced superconducting gap, while utilising the shadow masking technique~\cite{Carrad2020}. The absence of thin film formation was likely a consequence of low nucleation probability on the surface as well as high interface energy/strain to InAs. Growth of relatively thick In films ($>20$~nm) with a very low substrate temperature $< 120$~K and high flux may produce a large grain heteroepitaxial match to InAs.\cite{Ewert1976} We further speculate that a crystalline thin film may be obtained by the introduction of an intermediate layer (e.g Al) to increase the probability of forming nucleation sites on the surface or to mediate strain. In addition, a short flux of Sb before the In deposition may improve the wetting probability due to the surfactant effect, where Sb atoms on the surface influence the surface energy reducing the adatom average displacement \cite{Yang2014}. Alloying In (e.g. with tin, InSn is a $T_\mathrm{C}=6$ K superconductor~\cite{Mochiku2019}) could be an alternative way of promoting the surface wetting properties. Since amorphous In thin films also are superconducting ($T_C\sim 4.5$ K)~\cite{granqvist1975}, annealing the In grains to obtain an amorphous thin film could be another interesting approach, since amorphous films are capable of inducing hard gap superconductivity if the semiconductor/superconductor interface is pristine and the superconductor itself has a hard gap.~\cite{Carrad2020}

\section{Methods}

\subsection{Hybrid NW growth}
InAs NWs, $5-8~\mu$m-long, were grown on shadow epitaxy substrates \emph{via} the Au catalyst-assisted vapour-liquid-solid mechanism in a solid-source Varian GEN-II MBE system following the two-step protocol outlined in Ref.~\citenum{Carrad2020}. Such growth has been shown to produce flat NW facets, optimal for subsequent superconductor deposition.~\cite{Carrad2020,Kanne2020} Following NW growth, growths 2 and 3 were transferred \emph{via} UHV to an e-beam evaporation metal deposition chamber, where In deposition was performed at $T_{\mathrm{\mathrm{S}}} \sim -150^\circ$C with a tilt angle $\theta = 10^\circ$ along the $[11\bar{2}]$ direction for a 3-facet shell with nominal thickness 40~nm. The nominal deposition rates were 0.3~Å/s (growth 2) and 3~Å/s (growth 3). A third sample, growth 1, remained inside the growth chamber overnight to be cooled by the surrounding liquid nitrogen-cooled cryo-shield to $T_{\mathrm{\mathrm{S}}} \sim -37^\circ$C. Thereafter, In was deposited using an In effusion cell at $900^\circ$C for 4 minutes.

\subsection{Electron microscopy}
A JEOL 7800F scanning electron microscope with acceleration voltage $V_{acc} = 10-15$~kV was used to obtain Figs.\ref{fig:In_1}~a-g, and Fig.~\ref{fig:In_3}a. Prior to TEM characterisation, a micromanipulator under an optical microscope was used to place NWs on a carbon membrane grid. The micrographs were obtained using either a FEI Titan Analytical 80-300ST featuring a monochromator and $V_{acc} = 300$~kV (Figures 2b,c) or a JEOL 3000F with $V_{acc} = 300$~kV (Figure 2(a)). Double junction NWs (like those shown in Fig.~\ref{fig:In_1}e-g) were used for TEM analysis, and alignment performed using the bare InAs segment to ensure the beam was parallel to the InAs $\left\lbrace 1\bar{1}00\right\rbrace$ facets ($\left\langle2\bar{1}\bar{1}0\right\rangle$ zone-axes). The isolated In segment between the two bare InAs junctions was used for imaging due to an increased likelihood of finding isolated grains for analysis compared to other segments of the NW, where multiple grains were likely to be in the beam path.

\subsection{Device fabrication and measurement}
To fabricate devices, NWs were placed on a pre-patterned substrate using a micromanipulator. The ohmic contacts and side gates were patterned using electron beam lithography and e-beam evaporation of dual-layer Ti/Au. Low resistance contact to the InAs was obtained by Ar$^+$-ion milling immediately prior to contact deposition. Electron transport measurements were conducted using phase-sensitive lock-in detection at low frequency ($\sim 200$~Hz), in a dilution cryostat with base temperature $\sim 20$~mK.

\begin{acknowledgement}
This work was funded by the Innovation Fund Denmark's Quantum Innovation Center (Qubiz), the Carlsberg Foundation, the Niels Bohr Institute, the Villum Foundation (00013157), Microsoft Station Q and European Union’s Horizon 2020 research and innovation programme FETOpen grant no. 828948 (AndQC) and QuantERA project no. 127900 (SuperTOP). The Center for Quantum Devices is supported by the Danish National Research Foundation.

The authors thank Martin Aagesen, Claus B. S\o rensen and Shivendra Upadhyay for technical assistance and insightful discussions.
\end{acknowledgement}
\begin{suppinfo}
Supplementary information containing additional selected area electron diffraction analysis, structural analysis and electron transport data is found following this manuscript.
\end{suppinfo}

\section*{Data availability}
Full data sets for all figures, TEM images and electron transport data are available online at https://erda.ku.dk/archives/12007f3199be92dfa51477e7a265a320/published-archive.html

\bibliography{In_InAs.bib}

\end{document}

\renewcommand{\thefigure}{S\arabic{figure}}
\renewcommand{\thetable}{S\arabic{table}}

\section{Selected area electron diffraction analysis}

\begin{figure}[!htp]
\centering
\includegraphics[width=\textwidth]{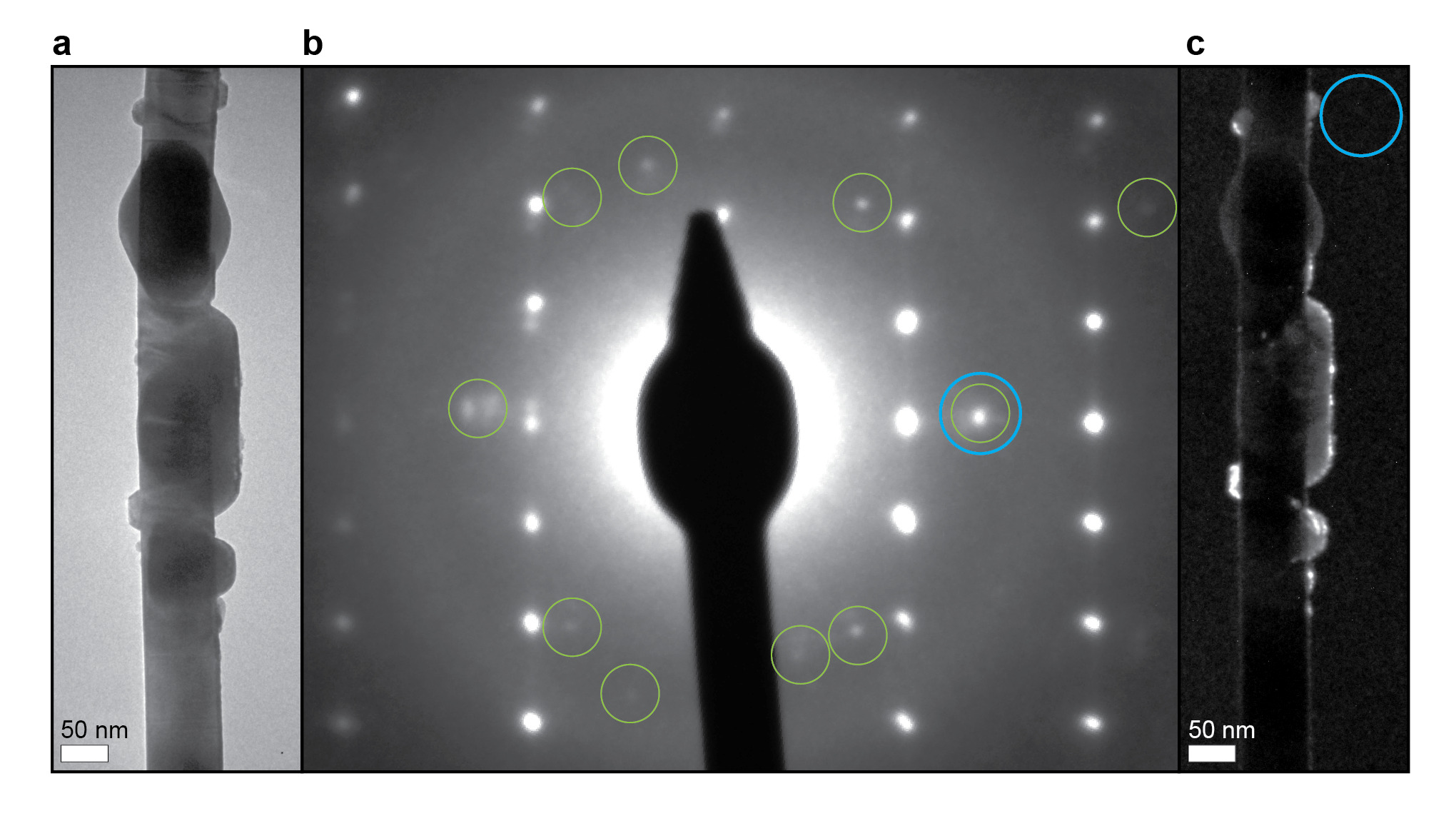}
\caption{\label{fig:In_Supp1}(a) TEM of an In/InAs hybrid NW shadow mask patterned by two SiOx bridges (similar to that in Fig. 1e-g in the main article). The InAs NW featured $\sim 12$ In grains of various sizes between the shadowed segments. (b) SAED signal associated with the image in (a). The rectangular peak grid arises from the InAs WZ crystal aligned parallel to $\{1\bar{1}00\}$ facet planes. Additional peaks (green circles) arose from the In grains. (c) Dark field signal from placing the objective aperture around In peaks aligned to the facet planes (positioning the aperture at the blue circle in (b)).}
\end{figure}

Transmission electron microscope (TEM), selected area electron diffraction (SAED), and dark field (DF) images from a double junction In/InAs NW are shown in Fig.~\ref{fig:In_Supp1}. The displayed NW region in Fig.~\ref{fig:In_Supp1}\textbf{a}, found between two shadow junctions, contained $\sim$~12 In grains of various sizes, from a projected grain area of $\sim$150~nm$^2$ up to $\sim$25 000~nm$^2$. The associated SAED signal (Fig.~\ref{fig:In_Supp1}\textbf{b}) shows that the InAs NW was aligned parallel to a $\langle2\bar{1}\bar{1}0\rangle$ zone-axis. Additional peaks (marked by green circles) arose from diffraction in the In crystals. Interestingly, an intense pair of In Bragg peaks aligned with the InAs $\{1\bar{1}00\}$ facet planes are seen (e.g. within blue circle). By placing the SAED aperture at the position of the blue circle in Fig.~\ref{fig:In_Supp1}\textbf{b}, the dark field (DF) signal associated with this Bragg peak was obtained (Fig.~\ref{fig:In_Supp1}\textbf{c}). The DF signal revealed the selected Bragg peaks to be common for all the In grains interfacing the aligned facets. All grains in that region therefore featured crystal planes that were parallel to the NW facet surface, in agreement with the anisotropic grain surfaces shown in the high resolution TEM images in Fig~2 of the main text.

\newpage
\section{Supplementary structural analysis}
In main text, we state the plane indices and orientations for the two In grains shown in Figs~2b,c. To make this plane assignment, we compared the measured interplanar distances and angles to the known values of various low-index candidate planes in bulk In. Bulk In crystallises in the tetragonal body-centred (TBC) phase (space group $I4/mmm$) with lattice parameters $a=0.3253$~nm and $c=0.4947$~nm~\cite{Haynes2016,Moshopoulou2006}. The interplanar distances, $d^{bulk}_{(hkl)}$, between various low-index TBC planes are given in Table~\ref{tab:InAsIn_TBC} and were calculated from~\cite{edington1975} 

\begin{equation}
	\dfrac{1}{d_{(hkl)}^2}= \dfrac{1}{a^2}(h^2+k^2)+\dfrac{1}{c^2}l^2
	\label{eq:TBC_dis}
\end{equation}

Similarly, the angles, $\phi^{bulk}$, between two planes, ($h_1k_1l_1$) and ($h_2k_2l_2$), are given in Table~\ref{tab:InAsIn_TBCangles} and were calculated using~\cite{edington1975}

\begin{equation}
	\cos \phi = \dfrac{\dfrac{1}{a^2}(h_1 h_2 +k_1k_2)+\dfrac{1}{c^2}l_1l_2}{\sqrt{\left[\left\lbrace\dfrac{1}{a^2}(h_1^2+k_1^2)+\dfrac{1}{c^2}l_1^2\right\rbrace\left\lbrace\dfrac{1}{a^2}(h_2^2+k_2^2)+\dfrac{1}{c^2}l_2^2\right\rbrace\right]}} \\
	\label{eq:TBC_angle}
\end{equation}

\begin{table}[ht]
\centering
\begin{tabular}{|l|r|r|r|r|r|r|}
\hline
TBC Planes		 & $d_{\{100\}}$ (Å) & $d_{\{001\}}$ (Å) 	  & $d_{\{101\}}$ (Å) & $d_{\{110\}}$ (Å) & $d_{\{111\}}$ (Å)  & $d_{\{112\}}$ (Å)   \\ \hline
			 &  3.253			 &  4.947				  &	2.718		      & 2.300			  & 2.085			   & 1.684	\\ \hline 
\end{tabular}
\caption{Interplanar distances, $d^{bulk}_{(hkl)}$, in the tetragonal body-centred (TBC) In phase. Calculated using Eq. \ref{eq:TBC_dis} and the lattice parameters found in Ref.~\citenum{Moshopoulou2006}.}
\label{tab:InAsIn_TBC}
\end{table}

\begin{table}[ht]
\centering
\begin{tabular}{|l|r|r|r|r|r|r|}
\hline
TBC angles $\phi^{bulk}$ ($^\circ$) & $\{100\}$& $\{001\}$& $\{101\}$ 	 	 	   & $\{110\}$  		   & $\{111\}$  & $\{112\}$    \\ \hline
		  $\{100\}$/$\{010\}$	 &  90.0	  &  90.0	  &	 33.3	 			   & 45.0				   & 50.1	    & 58.8		  	\\ \hline
					$\{001\}$	 &  90.0	  &  90.0   &	 56.7				   & 90.0	 			   & 65.1  	    & 47.1		  	\\ \hline
		  $\{101\}$/$\{011\}$	 &  33.3	  &  56.7	  &	 72.4	  	           & 53.8	               & 39.9	    & 36.2		  	\\ \hline
					$\{110\}$	 &  45.0	  &  90.0	  &	 53.8	               & 90		  			   & 24.9	    & 42.9/90   	\\ \hline
					$\{111\}$	 &  50.1	  &  65.1	  &	 39.9				   & 24.9				   & 79.8/49.9	& 18.0		  	\\ \hline
					$\{112\}$	 &  58.8      &  47.1	  &	 36.2		     	   & 42.9				   & 18.0 	    & 85.8		  	\\ \hline
\end{tabular}
\caption{Angles between families of planes in TBC In, $\phi^{bulk}$. Calculated using Eq. \ref{eq:TBC_angle} and the lattice parameters found in Ref.~\citenum{Moshopoulou2006}. When calculating the angle between planes from the same family, different planes within the family were used.}
\label{tab:InAsIn_TBCangles}
\end{table}

\begin{figure}[!htp]
\centering
\includegraphics[width=\textwidth]{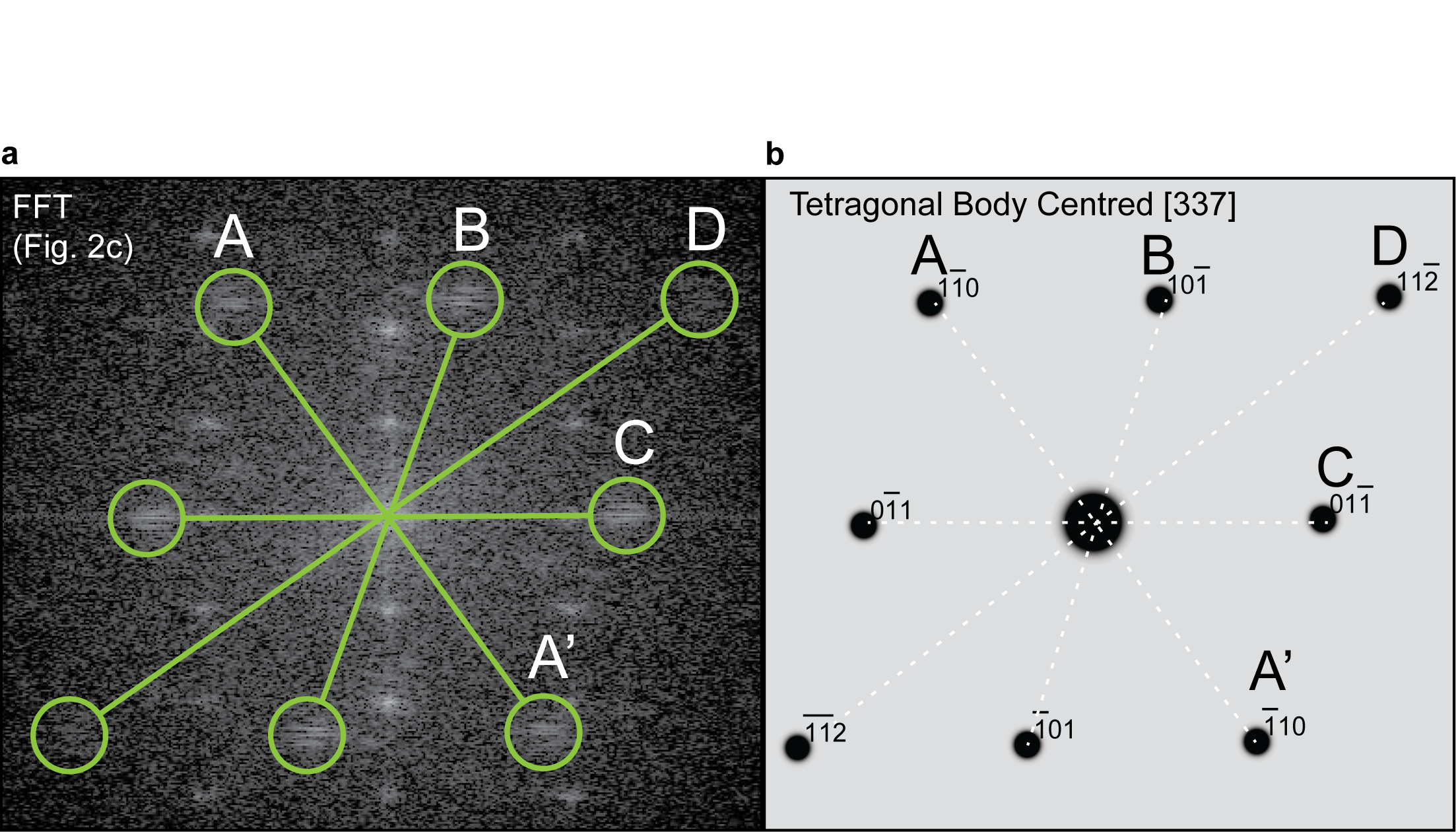}
\caption{\label{fig:In_Supp2} Labeling of the intensity peaks in FFT signal from (a) TEM micrograph of main text figure 2c. (b) Simulated electron diffraction signal from TBC crystal oriented along the [337] zone-axis.}
\end{figure}

Having established a list of potential candidate planes, we measured the interplanar distances, $d_{(hkl)}$, and angles between planes, $\phi$, using the real space HRTEM images and fast Fourier transforms (FFT), respectively, shown in main text Figs~2b,c. An example of the reference names used for the intensity peaks in the FFT (letters A-D) are shown in Fig. \ref{fig:In_Supp2}a, along with how they matched the model bulk TBC crystal (Fig. \ref{fig:In_Supp2}b). The measured values are compiled in Tables~\ref{Tab:In_distance} and \ref{Tab:In_angle}, along with calculated $d^{bulk}_{(hkl)}$ and $\phi^{bulk}$ from low-index bulk In planes that most closely match the measured values. The extremely good agreement gives us a high degree of confidence in the plane assignment stated in the main text. We used the same procedure to characterize the crystal orientation of all seventeen grains reported.

\begin{table}[ht]
\centering
\begin{tabular}{llll} \hline
\multicolumn{1}{|l|}{TEM Interplanar dist.}                                & \multicolumn{1}{|l|}{Avg. measured $d$} &  \multicolumn{1}{|l|}{Bulk In interplanar dist.}&   \multicolumn{1}{|l|}{Calc. $d^{bulk}_{(hkl)}$}         \\ \hline
\multicolumn{1}{|l|}{$d_A$}                                                     & \multicolumn{1}{l|}{2.3 Å $\pm$ 0.2 Å} & \multicolumn{1}{l|}{$d_{(110)}$}                    & \multicolumn{1}{l|}{2.30 Å} \\ \hline
\multicolumn{1}{|l|}{$d_B$}                                                     & \multicolumn{1}{l|}{2.8 Å$\pm$ 0.3 Å} & \multicolumn{1}{l|}{$d_{(101)}$}                      & \multicolumn{1}{l|}{2.72 Å} \\ \hline
\multicolumn{1}{|l|}{$d_C$}                                                     & \multicolumn{1}{l|}{2.8 Å$\pm$ 0.3 Å} & \multicolumn{1}{l|}{$d_{(101)}$}                      & \multicolumn{1}{l|}{2.72 Å} \\ \hline
\multicolumn{1}{|l|}{$d_D$}                                                     & \multicolumn{1}{l|}{1.7 Å$\pm$ 0.2 Å} & \multicolumn{1}{l|}{$d_{(112)}$}                      & \multicolumn{1}{l|}{1.68 Å} \\ \hline
\end{tabular}
\caption{Interplanar distance measurements, $d_{(hkl)}$, from HRTEM micrographs in Figs.~2b,c in the main article, compared to the bulk interplanar distances for closely matched low-index planes, $d^{bulk}_{(hkl)}$\cite{Moshopoulou2006}. All measurements were based on averaging 6-10 planes. The propagated uncertainty from calibration and measurement was $10\%$}
\label{Tab:In_distance}
\end{table}

\begin{table}[ht]
\centering
\begin{tabular}{llll}
\hline
\multicolumn{1}{|l|}{FFT angle} 								     			& \multicolumn{1}{|l|}{Avg. measured $\phi$} & \multicolumn{1}{|l|}{Bulk In angle} 			    & \multicolumn{1}{|l|}{Calc. $\phi^{bulk}$} \\ \hline
\multicolumn{1}{|l|}{$\measuredangle AOB \land \measuredangle A'OB$} 			& \multicolumn{1}{l|}{54.1$^\circ$ $\pm$ 3$^\circ$}   & \multicolumn{1}{l|}{$\angle \{110\}O\{101\}$} & \multicolumn{1}{l|}{53.8$^\circ$} \\ \hline
\multicolumn{1}{|l|}{$\measuredangle BOC$}                        			    & \multicolumn{1}{l|}{72.3$^\circ$ $\pm$ 3$^\circ$}   & \multicolumn{1}{l|}{$\angle \{101\}O\{011\}$} & \multicolumn{1}{l|}{72.4$^\circ$} \\ \hline
\multicolumn{1}{|l|}{$\measuredangle AOC \land \measuredangle A'OC$} 			& \multicolumn{1}{l|}{54.3$^\circ$ $\pm$ 3$^\circ$}   & \multicolumn{1}{l|}{$\angle \{110\}O\{101\}$} & \multicolumn{1}{l|}{53.8$^\circ$} \\ \hline
\end{tabular}
\caption{Interplanar angle, $\phi$ measurements from FFT signals of main text Figs.~2b,c, compared to caluculated bulk values $\phi^{bulk}$\cite{Moshopoulou2006}. From symmetry, $\measuredangle AOB$ ($\measuredangle AOC$) is equivalent to $\measuredangle A'OB$ ($\measuredangle A'OC$); hence they share a table row. $A'$ denotes the opposite reflection of $A$, introduced here to clarify the reference points for the angle measurements. The uncertainty for $\phi$ was calculated from the FFT peak full width at half maximum (FWHM).}
\label{Tab:In_angle}
\end{table}

\newpage
\section{Bi-crystal domain residual mismatch}
In our In/InAs hybrids, we observed crystal plane alignment between low-index In planes and the axial and transverse planes of the crystalline InAs NWs.
Bi-crystal domain residual mismatch -- the relative length difference between domains of single or multiple interplanar distances in crystals -- can be used to assess the potential for epitaxy and the anticipated strain between crystals.\cite{Krogstrup2015,Jelver2017,Kanne2020}. 

The bulk bi-crystal residual mismatch $\epsilon$ between $N$ In ($hkl$) planes and $M$ InAs ($hkl$) planes is calculated as

\begin{equation} \label{eq:bcm}
    \epsilon = \dfrac{N d_{\mathrm{In}}^{hkl}-M d_{\mathrm{InAs}}^{hkl}}{Md_{\mathrm{In}}^{hkl}}\cdot 100
\end{equation}

where $d_{\mathrm{In}}^{hkl}$ and $d_{\mathrm{InAs}}^{hkl}$ are the interplanar distances in In and InAs, respectively.
The theoretical bi-crystal interfacial matches between low-index TBC In interplanar distances and the interplanar distances of axial InAs (0002) and transverse InAs ($2\bar{1}\bar{1}0$) planes with the smallest domain and the lowest residual mismatches are shown in Table~\ref{tab:In_mismatch}. A lower $\epsilon$ value, along with fewer planes in either material is an indication of a more likely observed match as it indicates lower stress in the bi-crystal interface. 

\begin{table}[ht]
\begin{tabular}{|l|l|l|}
\hline
Planes      & Axial $\epsilon$ to InAs (0002) (\#In:\#InAs) & Transverse $\epsilon$ to InAs ($2\bar{1}\bar{1}0$) (\#In:\#InAs) \\ \hline
(100) & -7.4\% (1:1)               & 1.5\% (2:3)        	\\ \hline
(001)       & -6.1\% (2:3),5.6\% (3:4)   & -7.4\% (2:5), -0.8\% (3:7) \\ \hline
(111)       & -1.03\% (5:3)              & -2.4\% (1:1)   \\ \hline
(110)       & -1.8\% (3:2)               & 7.6\% (1:1)           \\ \hline
(101) & 3.2\% (4:3), -3.3\% (5:4)  & -4.6\% (3:4), 1.7\% (4:5)    \\ \hline
(112)       & -4.1\% (2:1)      & 5.1\% (4:3), -1.5\% (5:4)   \\ \hline
(211)       & -0.7\% (5:2)               & -2.1\% (3:2) \\
\hline
(337)       & -1.5\% (4:1)               & 3\% (4:1) \\ \hline  
\end{tabular}
\caption{Bi-crystal mismatch $\epsilon$ between In planes and InAs axial and transverse planes. The values in the table were based on 16 calculated domain mismatch tables spanning a wide range of domain sizes. Here, only the smallest mismatches of few plane (\#In:\#InAs) domains are shown.}
\label{tab:In_mismatch}
\end{table}

As presented in the main text, the observed In crystal planes that was found to align along the axial (0002) and transverse ($2\bar{1}\bar{1}0)$) InAs planes were ($11\bar{2}$) and (337), respectively. With a $-4.1\%$ mismatch between two In ($11\bar{2}$) planes and one (0002) InAs plane and a $3\%$ mismatch between four In (337) planes and one ($2\bar{1}\bar{1}0$) InAs plane, the observed bi-crystal alignment does not represent the smallest bi-crystal domains available. Compared to surface energy minimization, interface strain minimization by domain matching was thus considered a less dominant driving factor for the crystal grain growth.

\newpage
\section{Additional Transport Data}

\begin{figure}[!t]
\centering
\includegraphics[width=15cm]{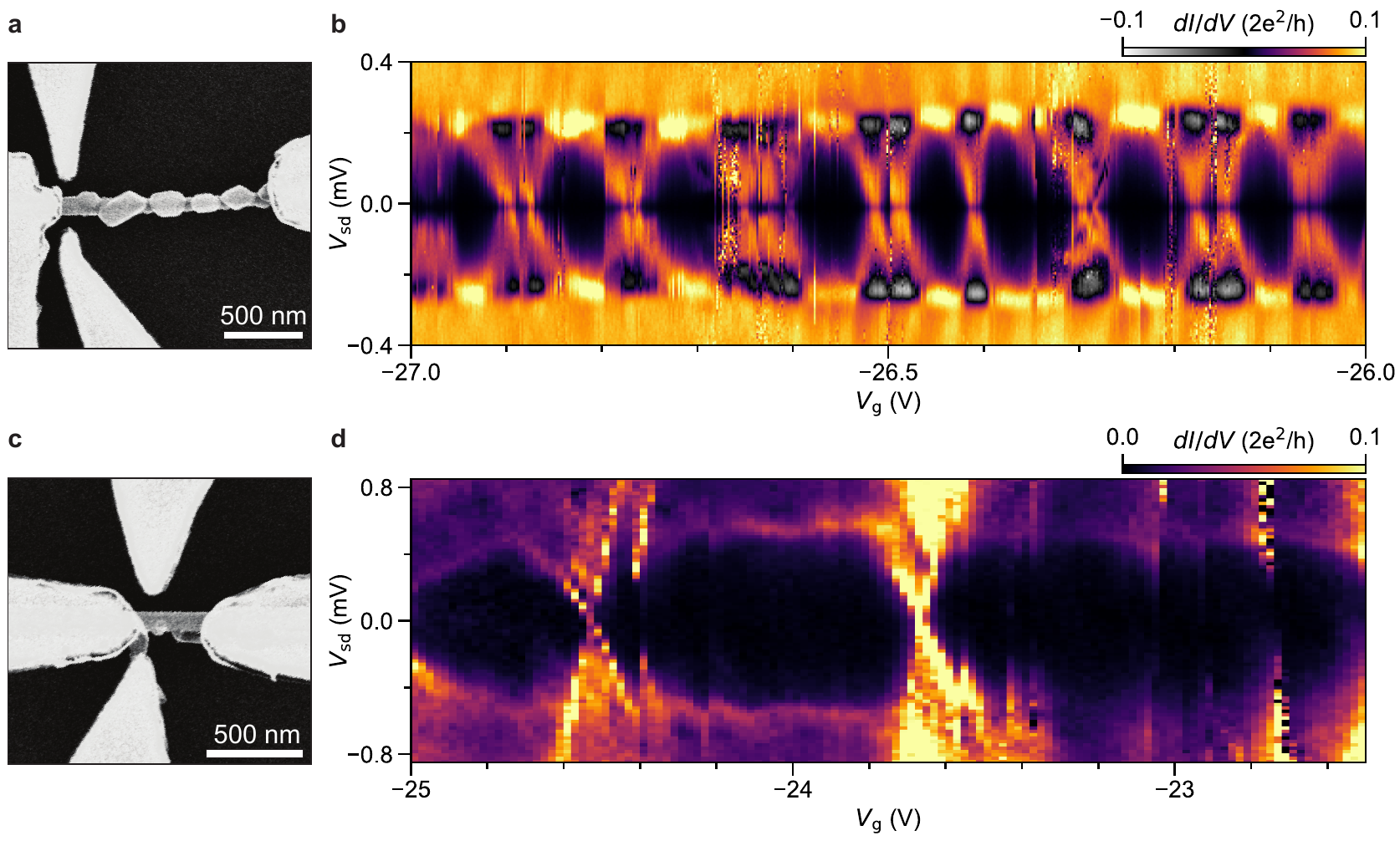}
\caption{(a/c) SEMs and (b/d) bias spectroscopy data of two additional In/InAs hybrid devices. Spectroscopy data acquired at $T=16$~mK and $B=0$~T.}
\label{fig:S3}
\end{figure}

\begin{figure}[!ht]
\centering
\includegraphics[width=16cm]{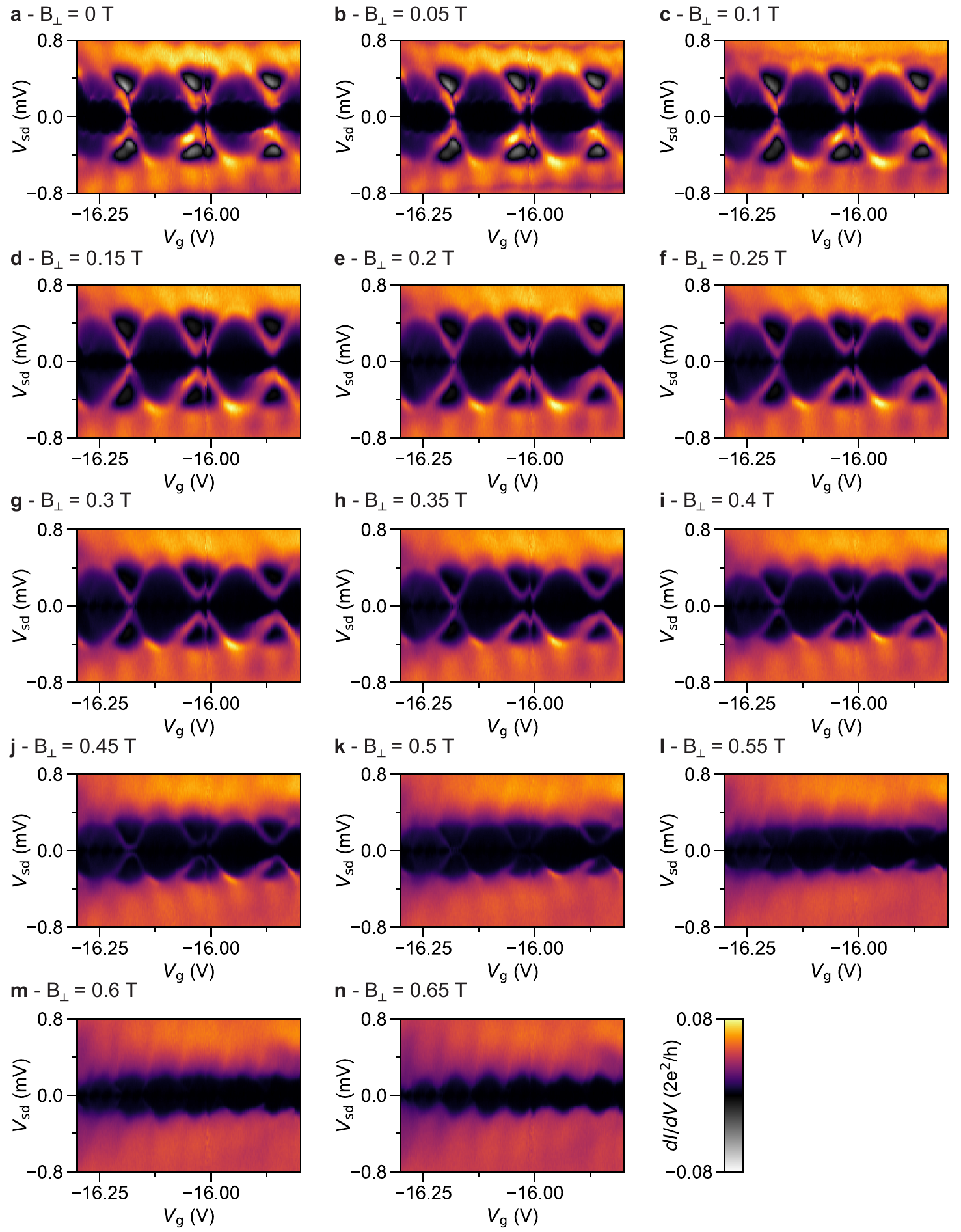}
\caption{Bias spectroscopy of the device shown in the main text at various fixed perpendicular magnetic field values $B_
\perp$.}
\label{fig:S4}
\end{figure}

In main text Fig.~3 we characterized an In/InAs hybrid device which acted as a superconducting island. This device showed 2e-periodic charging at low field, temperature and source-drain voltage $V_\mathrm{sd}$. In Supplementary Fig.~\ref{fig:S3} we show scanning electron micrographs and bias spectroscopy data from two additional devices. The first device (Fig.~\ref{fig:S3}a) featured a number of In islands between source and drain. Bias spectroscopy for this device is shown in Fig. \ref{fig:S3}b. The device properties exhibit instability as a function of $V_\mathrm{g}$, with various `switching' events causing features to be cut-off, or repeated along the x-axis. Such uncontrolled switching typically arises from changing charge states of nearby impurities. For this device, the additional islands along the nanowire length may have provided an additional switching source, since devices fabricated by shadow lithography are otherwise generally very stable.\cite{Carrad2020} Nevertheless, the key features of Cooper pair tunneling are observable, namely, regions of negative differential conductance below the gap edge due to quasiparticle poisoning, and a doubling of the resonance frequency in $V_\mathrm{g}$ above the gap.\cite{HergenrotherPRL94, AlbrechtNature16, ShenNatComm18, Kanne2020} The presence of 2$e$ charging again confirms the quality of the hybrid materials and interface.

The second device (Fig. \ref{fig:S3}c) acted as a `normal-superconductor' tunnel spectroscopy device at negative $V_\mathrm{g}$. Negative $V_\mathrm{g}$ induces a barrier in the conduction band profile of the InAs between the superconducting In, and normal metal Ti/Au source contact. Transport through this barrier -- which may be a quantum point contact or quantum dot -- produces a $dI/dV$ spectrum proportional to the density of states in the hybrid In/InAs hybrid.\cite{Mourik2012b, Chang2015} Fig.~\ref{fig:S3}d shows that as $V_\mathrm{g}$ is varied, diagonal lines crossing $V_\mathrm{sd}=0$ at, e.g., $V_\mathrm{sd}\sim -24.5 V, -23.65 V$ suggest the barrier took the form of a quantum dot. The horizontal $dI/dV$ peaks which remain at relatively constant $V_\mathrm{sd}\sim\pm 0.5$~mV with varying $V_\mathrm{g}$ can be attributed to the superconducting gap of In. 

In Supplementary Fig.~\ref{fig:S4} we show bias spectroscopy plots at various fixed magnetic fields $B_\perp$ for the device presented in main text Fig.~3. At $B_\perp =0$, 2e(1e)-periodic charging is observed below(above) the superconducting gap. As increased $B_\perp$ reduces the value of the gap, the 1e charging moves to lower $V_\mathrm{sd}$, until it takes place around zero bias. This behaviour matches that seen in Fig.~3f of the main text.

\bibliography{InInAsreferences.bib}